\documentstyle[12pt,epsfig,amssymb]{article}
\begin{document}
\def\be{\begin{equation}}
\def\ee{\end{equation}}
                         \def\bearr{\begin{eqnarray}}
                         \def\eearr{\end{eqnarray}}
\def\benum{\begin{enumerate}}
\def\eenum{\end{enumerate}}
\def\bitem{\begin{itemize}}
\def\eitem{\end{itemize}}
                         \def\eg{ {\em e.g.}~}
                         \def\etal{ {\em et al.}~}
                         \def\ie{ {\em i.e.}~}
                         \def\viz{ {\em viz.}~}

\def\go{\rightarrow}
\def\goes{\longrightarrow}
\def\hrar{\hookrightarrow}
\def\bul{\bullet}
\def\eplem{\mbox{$e^+e^-$}}
\def\gamgam{\mbox{$\gamma \gamma$}}
\def\gamp{\mbox{$\gamma p$}}
\def\rts{\mbox{$\sqrt{s}$}}
\def\ptmin{\mbox{$p_{\rm tmin}$}}
\def\siggmjet{\mbox{$\sigma (\gamma \gamma \rightarrow {\rm jets})$}}
\def\lsim{\:\raisebox{-0.5ex}{$\stackrel{\textstyle<}{\sim}$}\:}
\def\gsim{\:\raisebox{-0.5ex}{$\stackrel{\textstyle>}{\sim}$}\:}

\begin{flushright}
LNF-00/024(P)\\
IISc-CTS/15/00\\
hep-ph/0010104
\end{flushright}

\begin{center}

{\large\bf    
        Hadronic  cross-sections in \gamgam\ processes and the
        next linear  Collider } \\

\vskip 25pt

{\bf                        Rohini M. Godbole } \\ 

{\footnotesize\rm 
                      Centre for Theoretical Studies, 
                    Indian Institute of Science, Bangalore 560 012, India. \\ 
                     E-mail: rohini@cts.iisc.ernet.in  } \\ 

\bigskip

{\bf                       G. Pancheri } \\ 

{\footnotesize\rm 
                    Laboratori Nazionali di Frascati dell'INFN,  
                     Via E. Fermi 40, I 00044, Frascati, Italy. \\ 
                     E-mail: Giulia.Pancheri@lnf.infn.it  } \\

\vskip 30pt

{\bf                             Abstract 
}

\end{center}

\begin{quotation}
\noindent
In this note we address the issue of theoretical
estimates of the hadronic cross-sections for \gamgam\ processes.
We compare the predictions of the minijet model with data
as well as other models, highlighting the band of 
uncertainties in the theoretical predictions as well as those 
in the final values of the $\sigma (\gamgam \go {\rm hadrons}) $ 
extracted from the data. We find that the rise of
$\sigma_{\gamma \gamma}^{\rm tot}$ with energy shown in the latest
\gamgam\ data is in tune with the faster rise expected in the Eikonal
Minijet Models (EMM). We present an estimate of the accuracy with which this 
cross-section needs to be measured, in order 
 to distinguish between the different 
theoretical models which try to `explain ' the rise of total 
cross-sections with energy. We find that the precision of measurement
required to distinguish the EMM type models from the  proton-like 
models, for $ 300 < \sqrt{s}_{\gamgam} < 500 $ GeV, is 
$\lsim 20 \% $,  whereas to distinguish between various proton-like
models or among the different parametrizations of the EMM, a precision  of
$\lsim 8-9 \% $ or $\lsim  6-7 \% $ respectively, is required.  
We also comment briefly on  the implications of these  predictions 
for hadronic backgrounds at the  next linear collider (NLC) to be run 
in the \gamgam\ mode or \eplem\ mode. 

\end{quotation}

\vskip 60pt
\newpage
\section{ Introduction}
The rising total cross-section in proton-proton collisions was a very 
early indication of QCD processes at work, reflecting the fact that the
increasing energy allows for deeper and deeper probe of the structure of 
the colliding particles\cite{therise} leading to  liberation of  
more and more constituents,  resulting in higher scattering
probability. The proton-proton and proton-antiproton cross-sections  are  
now  known experimentally to a very good precision. We do not
yet fully understand these cross-sections from first principles, but 
there are various models of hadronic
interactions whose parameters can be completely fixed by the data and
which then allow for good predictions of the total cross-section in the 
high energy region, certainly up to LHC energies. 
Thus, although not everything is calculable from first principles in QCD, 
there is no problem as far as predicting the total hadronic production at 
future accelerators is concerned. The situation is 
substantially  different for the photon induced processes.  
This renders the issue of measurement of  the total \gamgam\ cross-section 
at energies in the region 300-500 GeV range very important both from 
the theoretical point of view, as well as experimental. Indeed,the 
question  of hadron production in \gamgam\ collisions is
interesting from a point of view of achieving a good theoretical
understanding of the rise of the hadronic cross-sections with 
energy, in the framework of QCD or otherwise, 
as well as from a much more pragmatic viewpoint of being able to
estimate the hadronic backgrounds at the next linear colliders~\cite{lcbkgd}. 
There exist  two classes of  models which have been
suggested in the context of rise of the hadronic cross-sections in
$pp$ and $\bar p p$ processes. All of these `explain' the
rise for the $pp$ and $\bar p p$ case equally well but differ greatly
in their predictions for \gamgam\ collisions even at the modest
values of the \gamgam\ energies that are currently available.
HERA and LEP have opened the way to an entire new field in QCD, the study of
the hadronic interactions of the photon in terms of its  quark
and gluon content~\cite{review}.  Hadronic collisions show the beginning of
the rise to take place at centre of mass energies below 100 GeV 
but to determine the steepness of  the rise one needs points in the 300-500 GeV 
and even higher.  Thus, as in the case of hadron collisions, to  gain a
good theoretical understanding of the total cross-section for \gamgam\ 
processes, one needs much higher energies and better statistics than  the one
currently  available from LEP and HERA. 
At LEP phase space limits the c.m. energy of the \gamgam\ system to  
about 100 GeV, at HERA $\sqrt{s}_{\gamma p}$ is higher,
but then the presence of proton partly obscures the issue. Note
that, until five years ago, the available data for \gamgam\  processes did not 
yet show any rise, stopping short of $\sqrt{s} =  20$ GeV and with very 
large errors. L3 and OPAL data have drastically changed the situation. 
Presently, the \gamgam\  cross-section data  show a very clear rise, but
there are a number of theoretical and experimental issues, 
which only the Linear Collider(LC)  can clarify, by  reaching  higher energies
and statistics. In what follows we shall discuss the theoretical issues
which can only be resolved by  measurements of total cross-sections at 
higher and higher energies in the \gamgam\ system.  In this note we first 
discuss  predictions~\cite{emmus} for $\sigma (\gamgam\ \to {\rm hadrons})$, 
in the Eikonal Minijet Model (EMM), and then make a comparison with other 
models as well as the  presently available experimental data, contrasting 
the uncertainties in the model predictions with those in the values of 
$\sigma (\gamgam\ \to {\rm hadron})$ extracted from the existing 
two-photon data~\cite{OPAL,L3,L3osaka}.  
\section{QCD and total cross-sections}
QCD predicts the rise of total cross-sections through the increasing 
number of gluon-gluon collisions. For this, one can take the
approach of the BFKL equation, which resums all the QCD diagrams, using
a frozen value of $\alpha_s$ for the final evaluation, the underlying idea 
being that both the decrease, "Regge type" behaviour, as well as
the rise,  {\it i.e.} the "Pomeron" like behaviour are QCD effects.  
The calculation gives  imaginary part of all the summed diagrams
and hence the total cross-section. This method does not yet allow for a 
precision estimate.  Hence, while from a theoretical point of view this is
fundamentally the correct approach to follow,  at present it does not 
have much predictive power.  A different, more pragmatic approach is
the one of the EMM\cite{eikminijets}, which uses the 
eikonal approximation to calculate the elastic amplitude and hence 
the imaginary part, i.e.  the total cross-section, and the QCD jet 
cross-section drives  the rise with energy. Notice that
the eikonal approximation is indeed just an approximation and hence
has its inherent limitations.  The neglected terms can, in
principle,  play a non-negligible role. The interest in using the 
Mini-jet Model\cite{minijet} (we will discuss the 
details  of the model later) lies in the fact 
that the phenomenological triumphs of QCD are based
on the possibilty of predicting the jet cross-sections in terms of  
a set of basic parton processes and  parton densities parametrizing 
the parton distributions in the proton or in the photon. 
Hence, it is this aspect of QCD which can be put to trial in 
calculating the rise of total cross-sections using the EMM approach, {\it viz.}
the validity of our description of the scattering processes through 
the above two ingredients. Since we are dealing with a total cross-section, 
it is only the inclusive jet yield  given by 
\begin{equation}
 \sigma^{\rm jet}(s,p_{\rm tmin})=
 \int_{p_{\rm tmin}}
 d^2{\vec p_t} {{d\sigma_{jet}}\over{d^2{\vec p_t}}}.
\label{minjets}
\end{equation}
which enters into the calculation. Here ${{d\sigma_{jet}}/{d^2{\vec p_t}}}$,
is the differential jet cross-section calculated using QCD.
This result depends upon the minimum jet transverse momentum used
in the calculation  and  the choice of parton densities  for the hadron.
For the case of the photon the available parametrizations in the low-x region 
differ from each other substantially. 

\begin{figure}[htb]
\begin{center}
\epsfig{file=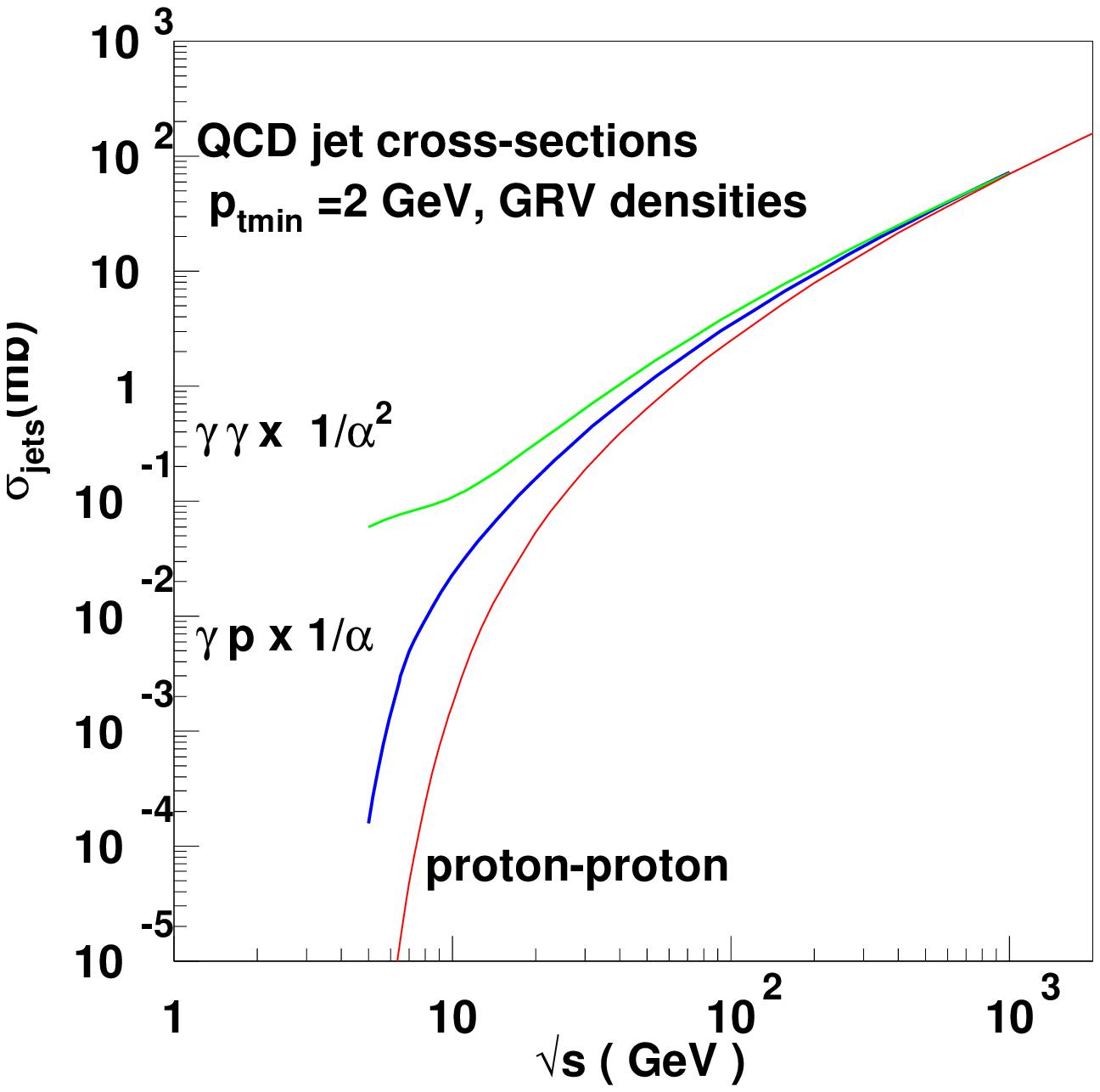,width=10cm}
\caption{Integrated jet cross-sections for $p_{\rm tmin}=2\ GeV$. }
\label{jets}
\end{center}
\end{figure}
In Fig.\ref{jets} we show the minijet 
cross-section defined by eq.~\ref{minjets} for three different types
of processes {\it viz.},   proton-proton, photon-proton  and photon-photon 
interactions.  Of course, in order to be able to compare all of them together, 
we have normalized them, by a multiplication with an appropriate power of 
$\alpha$, the fine structure constant. In this figure we have chosen 
$p_{\rm tmin}=2\ GeV$, a rather conservative value, based upon various 
phenomenological attempts at fitting $\sigma^{\rm tot}$ in different cases. 
In both the cases where photon is one of the colliding particles, 
two types of processes contribute  to the jet cross-sections:~the  
`resolved' processes, in which partons in the photon participate 
in the hard scattering process producing the jets  
and the `direct' processes in which a photon  participates 
directly  in the `hard' subprocess generating the final
state jets. This accounts for part of the difference between the three 
curves.
Another important difference is the markedly harder  $x$-dependence of 
the quark densities in the photon reflecting the hard nature of the
$q \bar q \gamma$ vertex, which gives rise to the perturbative part
of the photonic quark densities.  In the region where the gluon content is 
expected to dominate, the three curves superimpose as they should. 
In this figure we have chosen the  GRV\cite{GRV} densities for protons 
as well as for the  photons. However, the general features we mention 
are insensitive to our choice of the photonic densities.
The rise of the (mini)jet cross-sections with $\sqrt{s}$ is however 
too steep for any
fixed $p_{\rm tmin}$ value, and these cross-sections need to be 
incorporated in
a theoretical framework which preserves unitarity,  to allow for the 
calculation of a total cross-section.
The eikonal formulation provides a framework in which the minijet 
cross-sections  are unitarised via multiple scattering~\cite{treleani}.
Let us write down, very briefly, the formulation of the EMM for the 
$pp/ \bar p p$ case. The starting point is the eikonal formulation for the 
elastic scattering amplitude.
\begin{equation}
f(\theta)={{i k}\over{2\pi}} \int d^2{\vec b} e^{i{\vec q}\cdot{\vec b}}
[1-e^{i\chi(b,s)}]
\label{elsc}
\end{equation}
Along with the optical theorem this leads to the expression for
the total cross-sections
\bearr
\label{eik1}
\sigma^{\rm el}_{pp(\bar p)}&=&\int d^2{\vec b}
|1-e^{i\chi(b,s)}|^2 \\
\label{eik2}
\sigma^{\rm tot}_{pp(\bar p)}&=&2\int d^2{\vec b}
[1-e^{-\chi_I(b,s)}cos(\chi_R)] \\ 
\label{eik3}
\sigma^{\rm inel}_{pp(\bar p)}&=&\sigma^{\rm tot}-\sigma^{\rm el}=\int d^2{\vec b}
[1-e^{-2\chi_I(b,s)}]
\end{eqnarray}
The jet cross-section is a priori an inelastic part of the total cross-section, and thus is used as an input to the eikonalized formulation of 
the {\it inelastic} cross-section. 
The eikonal formulation introduces a new set of uncertainties in this 
calculation, {\it viz.} the impact parameter distribution in the colliding 
particles, a semi-classical concept, whose range of validity is limited to 
the large b-collisions.  The arbitrariness introduced by this function 
needs to be evaluated carefully.
It plays a role  not only in the low energy behaviour, but also, 
and very much so, in the rise. It is possible to relate this function to 
QCD processes, and attempts to calculate it have been partially 
successful\cite{bn}.  Of course, the rise with energy 
is only one of the phenomenological 
aspects of total cross-sections. Hence one should expect the 
function $\chi_I(b,s)$ to have both a soft part which is nonperturbative in
origin and cannot be calculated in this context, and a `hard' component which 
can be.
A useful working scheme is one in which one
breaks down the eikonal function in a number of building blocks, and then 
examines their phenomenological implications, varying the input 
parameters in each one of them. To be concrete, we write
\begin{equation}
2 \chi_I(b,s)=n_{\rm soft}(b,s)+n_{\rm PQCD}(b,s)
\label{chi}
\end{equation}
with 
\begin{equation}
n_{\rm soft}(b,s)=A_s(b,s)\sigma^{\rm soft}(s)
\end{equation}
and
\begin{equation}
n_{\rm PQCD}(b,s)=A_h(b,s)\sigma^{\rm jet}(s,p_{\rm tmin})
\end{equation}
Here $\sigma_{\rm jet}$ is given by eq.~\ref{minjets}.
This breakdown uses the assumption of factorization into
the transverse and longitudinal degrees of freedom, in other words between
the impact parameter distribution and the energy dependence. In this paper
we have fitted the data using  a further approximation: 
\begin{equation}
2 \chi_I(b,s)\equiv n(b,s)=A(b)[
\sigma^{\rm soft}(s)+
\sigma^{\rm jet}(s,p_{\rm tmin})
]
\label{numb}
\end{equation}
{\it viz.} assuming the same impact parameter distribution function
for soft and hard processes. In addition we have also
assumed independence of the overlap function on $s$. This approximation 
may be too rough, but, lacking a full theoretical description for 
this function, we prefer to study the energy behaviour only through the
hard cross-section, and modify this assumption only later.
Note that $\sigma_{jet}$  can be calculated completely in perturbative 
QCD once we have the knowledge of parton densities in the hadrons involved.
For the case of photon induced processes, the above formulation
needs to be generalised~\cite{ladinsky}to include the probability that 
photon behaves like a hadron in the collision. This generalisation has
been implemented in the earlier analyses of $\sigma^{\rm inel}_{\gamma p}$ 
~\cite{FLETCHER,SARC}. We denote  this probability by 
$P^{\rm had}_{\rm ab}$, which is unity for hadron-hadron (denoted by a,b)
processes, but of order $\alpha_{\rm em}$ or 
$\alpha_{\rm em}^2$ for processes where  one or both of the hadrons 
participating in the collisions are photons. Using Vector Meson 
Dominance~(VMD), and a running $\alpha_{\rm em}$,
this parameter varies from 1/250 at 
$\sqrt{s}=5$ GeV to 1/240 at $\sqrt{s}=300$ GeV. 
The generalisation involves putting the factor $P^{\rm had}_{\rm ab}$ 
on the R.H.S. in eqs.~\ref{eik2},~\ref{eik3} and dividing the second term 
in the square bracket in eq.~\ref{numb} by the same factor. Of course 
the $p$ 
in the subscripts will get changed to $\gamma$ depending on the number of 
photons involved in the initial state. The definition of
$\sigma^{\rm soft}$ in  eq.\ref{numb} is such that, even in the
photonic case, it is of hadronic size. Hence only the second term
in the square bracket of eq.~\ref{numb} gets the factor of 
$P_{\rm ab}^{\rm had}$. 
A simple way to understand the need for this factor~\cite{ladinsky} is 
to realise that the unitarisation in this formalism is achieved by 
multiple parton  interactions in a given scatter of hadrons and once 
the photon has `hadronised' itself, one should not be paying the price 
of $P_{\gamma}^{\rm had}$ for further multiparton scatters.
The eikonalised cross-sections depend only on a particular combination of
the hadronic factor $P^{\rm had}_{\rm ab}$ and the impact parameter 
distribution function $A_{ab}(b)$.   This, together with
the simple scaling properties of the eikonal formulation, allows for
an interesting graphical description of the b-distribution in the
three different cases of proton-proton, $\gamma\ $proton and 
photon-photon collisions.
On dimensional grounds, the function $A_{ab}(b)$ in general depends on two
scale parameters, describing respectively the matter distribution in the two
colliding particles a and b. Denoting such parameters by $k_a$ and $k_b$,  
where  $a,b=\gamma$ or $proton$, and assuming that the matter distribution 
in b-space factorizes into the Fourier transform of the electromagnetic  
form factor of the colliding particles, we have
\begin{equation}
A_{ab}(b)\equiv A(b;k_a,k_b)=
{{1}\over{(2\pi)^2}}
\int d^2 {\vec q} e^{iq\cdot b}
{\cal F}_a(q,k_a){\cal F}_b(q,k_b)
\end{equation}

\begin{equation}
\equiv {{1}\over{(2\pi)^2}}\int d^2 {\vec q} 
e^{iq\cdot b}
{\cal F}_a(q/k_a){\cal F}_b(q/k_b)
\end{equation}
The
general expression of the inelastic cross-section now reads
\begin{equation}
\label{sigin}
\sigma^{inel}_{ab}=P^{\rm had}_{\rm ab}\int d^2{\vec b}[1-e^{-n(b,s)}]
\end{equation}
To see the effect of $P^{\rm had}_{\rm ab}$ on the 
b-distribution, we scale out the factor $P^{\rm had}_{\rm ab}$, 
obtaining
\begin{equation}
\sigma^{inel}_{ab}=
\int d^2{\vec b}
[1-e^{-
{\bar A}(b;{\tilde k}_a,{\tilde k}_b)
[P^{\rm had}_{\rm ab}\sigma^{soft} + \sigma^{jet}]}]
\end{equation}
with 
${\bar A}(b;{\tilde k}_a,{\tilde k}_b)=A(b;{\tilde k}_a,{\tilde k}_b)
/P_{ab}^{\rm had}$
where ${\tilde k}_{a/b}=k_{a/b}/\sqrt{P^{\rm had}_{\rm ab}}$.
With this, all three total/inelastic cross-sections, $p\ p/{\bar p}$, 
$\gamma \ p$ and $\gamma \gamma$ can be obtained from the same expression,
with different overlap functions. 
\begin{figure}[htb]
\begin{center}
\epsfig{file=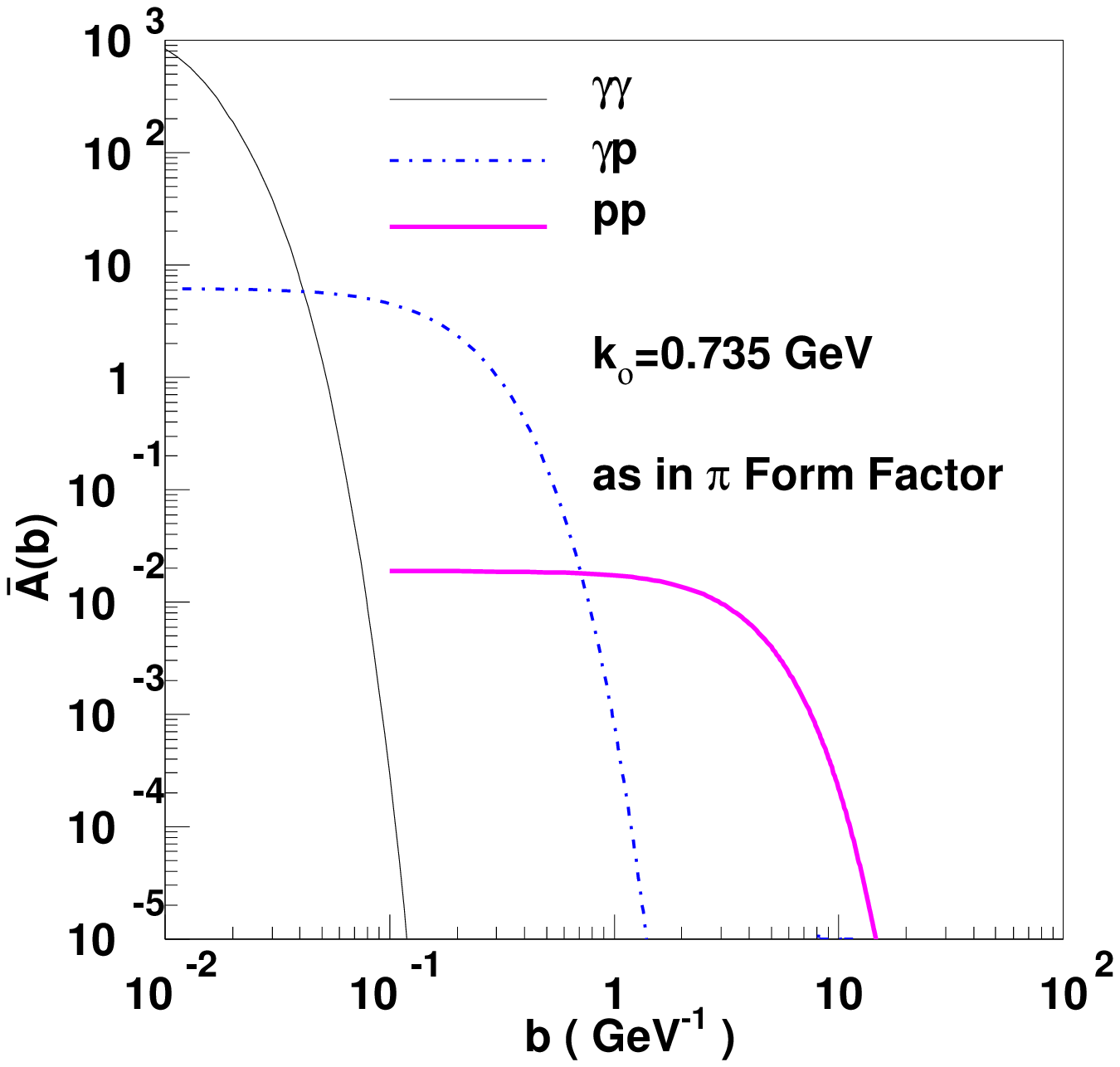,width=10cm}
\caption{The impact parameter distribution for protons and photons}
\label{aoball}
\end{center}
\end{figure}
To see the effects of these differences between photons and
protons~\cite{lundus}, we have plotted in Fig.\ref{aoball} the function
${\bar A}(b;{\tilde k}_a,{\tilde k}_b)$, for the case of 
$P^{\rm had}_{\gamma p}=
1/240$, using the usual dipole function for the proton form factor and the
pion form factor for the photon. This figure graphically emphasizes the 
difference in
spatial extension between photons and protons.

For  calculations, we revert to eq.(\ref{sigin}),
which, in the above eikonal formulation, very clearly defines
the  inelastic cross-section.
Making the
further approximation, well borne by experiment, that the real part of 
the eikonal $\chi_R(b,s)$ is zero, one is in principle ready to fit 
the data for the total photon-photon cross-sections, using
\begin{equation}
\label{sigtot}
\sigma^{tot}_{ab}=2 P^{\rm had}_{\rm ab}\int d^2{\vec b}[1-e^{-n(b,s)/2}]
\end{equation}
 The strategy adopted in \cite{emmus} consisted
in determining the various parameters, $\sigma^{\rm soft}, 
P_{\gamma p}^{\rm had},k_{\rm proton},k_{\gamma},
p_{\rm tmin}$ and the choice of photon densities, 
using the process $\gamma p $ and then, predict cross-sections 
for \gamgam. Next, for those parameters/functions 
which are non perturbative and need to be modified, we use factorisation;
{\it viz.} $P_{\gamma \gamma}^{\rm had} = [P_{\gamma p}^{\rm had}]^2$ 
and the Quark Parton Model, i.e.  $\sigma^{\rm soft}_{\gamma \gamma }
= {{2}\over{3}} \sigma^{\rm soft}_{\gamma p}$.
For the hard part of the eikonal, we keep 
the same photonic parton  densities, $p_{\rm tmin}$, and $k_\gamma$ 
as in the $\gamma p$ case. Note that the value of
$k_\gamma$ used in~\cite{emmus} corresponds to a different ansatz 
in the case of photonic partons than the one
mentioned in the discussion above.
In \cite{emmus} we have taken the matter distribution to be the 
Fourier Transform 
of the transverse momentum distribution. Technically it only means using the
experimentally measured value of $k_\gamma$, that
 is $ 0.66 \pm 0.22$~\cite{ZEUS}, 
instead of the value of $0.735$ used, for illustration, in Fig.~\ref{aoball}.
The correlated predictions of the EMM for $\gamma p$ and $\gamma \gamma$ 
have been discussed in \cite{photon99}.
The resulting curves for $\gamma \gamma$ total cross-section are shown in
\begin{figure}[htb]
\begin{center}
\epsfig{file=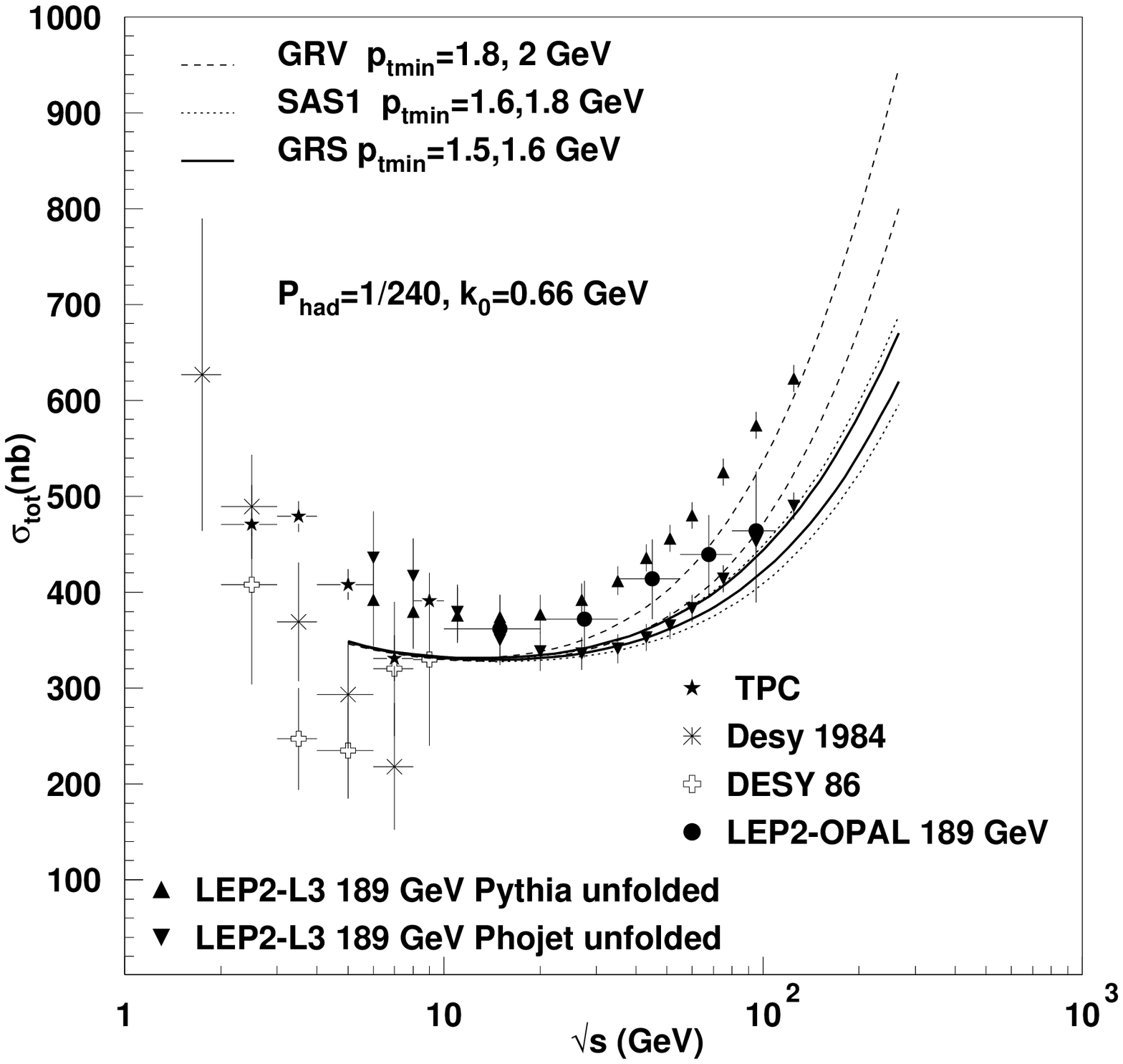,width=10cm}
\vspace{1cm}
\caption{Predictions of the EMM for different densities}
\label{figtwo}
\end{center}
\end{figure}
Fig.\ref{figtwo} for three different sets of parton densities~\cite{GRV,SAS,GRS}
and compared with data for $\sqrt{s} = 189$ GeV, 
from the L3\cite{L3} and OPAL~\cite{OPAL} Collaborations respectively.  
We have chosen a set of parameters which  
give compatible fits to both the HERA\cite{HERAZ,HERAH1} and the LEP 
data. It should be noted that all these curves give predictions which lie 
below the L3 data extracted with Pythia Monte Carlo~\cite{L3}.
 The most recent data analysis by the L3 Collaboration, inclusive of 
$202$ GeV data~\cite{L3osaka}, obtained
averaging between Pythia and Phojet, is now in full agreement with the OPAL 
data. At the same time, recently, 
new data for the $\gamma p$ cross-sections, extracted from Deep 
Inelastic Scattering have appeared\cite{DIS}, and new photoproduction 
data should be available soon. Thus complete reliance on the extrapolation 
from $\gamma p$ is not yet advisable. A 10\% change of the
parameters of the EMM prediction, for instance, can now give
a very good agreement with the
present data, as we show in 
 Fig.\ref{figmaria}, where, 
\begin{figure}[htb]
\begin{center}
\epsfig{file=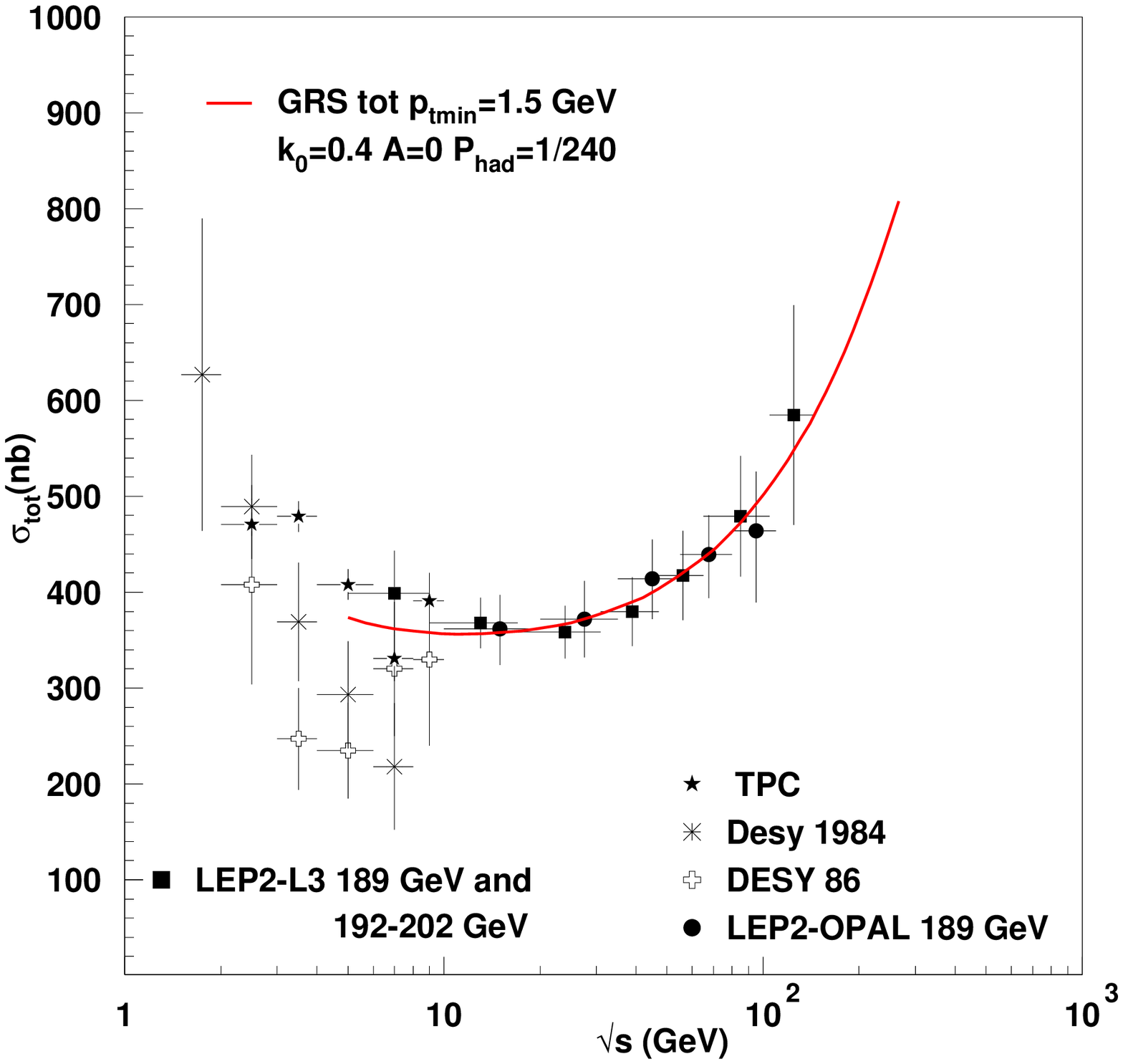,width=10cm}
\vspace{1cm}
\caption{EMM description of present $\gamma \gamma$ total
cross-section, with parameters extracted from $\gamma p$ with 
10\% changes as indicated in the text.} 
\label{figmaria}
\end{center}
\end{figure}
to obtain this figure,
we have followed the procedure described above, except that the 
value for the photon intrinsic transverse momentum $k_\gamma$ used is $0.4$ 
GeV which is lower than the $k_\gamma=0.66$ GeV, i.e.  the central value of 
$k_\gamma$ given by experiment~\cite{ZEUS} and which produces good fits of the
EMM predictions to the published photoproduction data on 
$\sigma^{\rm tot}_{\gamma p}$\cite{HERAZ,HERAH1}.
The other relevant parameters are
\begin{equation}
P^{\rm had}_{\gamma p} \equiv P_{\rm had} ={{1}\over{240}}\ ; \ \ \ \ 
\sigma^{soft}=\sigma_0+{{\cal B}\over{s}}\ ; \ \ 
\sigma_0=20.8\ mb \ ; \  {\cal B}=42.1 {{mb} {{\rm GeV}^2}} 
\end{equation}

All the minijet curves we have shown indicate a rather steep rise
of the total cross-section. There exist however other predictions, which 
one could call ``proton-like'' predictions, which reflect the validity of 
factorization for total cross-sections at present energies and in which 
the predicted rise at high energy is very different from the 
Eikonalised Minijet Model. 
This is the case of the Aspen Model\cite{aspen}, in which both the rise and 
the shape of the $b-$distribution are derived from the proton with simple
scaling properties. 
Although at {\it very } high energies this may not be true, at present 
energies this model satisfies the factorization hypothesis
$\sigma_{\rm nn}=\sigma_{\gamma p}^2/\sigma_{\gamma \gamma}$\cite{ttwu}. 
Another ``proton-like'' model is the Regge-Pomeron exchange model, in 
which factorization (at the residues) holds independently
for the low (Regge) energy and the high (Pomeron) energy term.
Most  of the ``proton-like'' models have the same high energy rise 
in all the three $pp$,$\gamma\ p$ and $\gamma \gamma$ cross-
sections, as typified by the Regge-Pomeron exchange model, where
$\sigma^{tot}\approx s^\epsilon$,
 with $\epsilon=0.08$\cite{DL}. There is a priori no justification in these 
models
for a change in  curvature from hadrons to photons, although data have
recently been parametrized with different values of the Pomeron
intercept 
parameter $\epsilon$.
On the other hand, the different rate at which $\sigma$ rises with energy
in the other  models has, obviously, both theoretical and 
experimental implications. The latter  because, as shown later, 
the predictions of different models can differ by a factor 2 or 3 
at the values of energy of interest to NLC and since these cross-sections 
enter in the calculations of photon-induced hadronic background,
the corresponding error in the prediction is then quite large. 
But even more important is the issue of arriving at a theoretical 
understanding of these differences and resolution as to which is 
theoretically more satisfactory and trustworthy.  
It is also worthwhile to determine whether the faster rise in minijet 
models is to be traced to the extrapolated low-x behaviour of the parton 
densities or is it inherent to the eikonal model.  An understanding
of how a different curvature can arise, in principle,  can be obtained 
from the EMM model. As shown in Fig.\ref{jets}, at high energy the 
minijet cross-sections, obtained from gluon-gluon scattering, all
rise with the same slope. However, their convolution with different impact 
parameter distributions changes the pattern.  To see this, we
\begin{figure}[htb]
\begin{center}
\epsfig{file=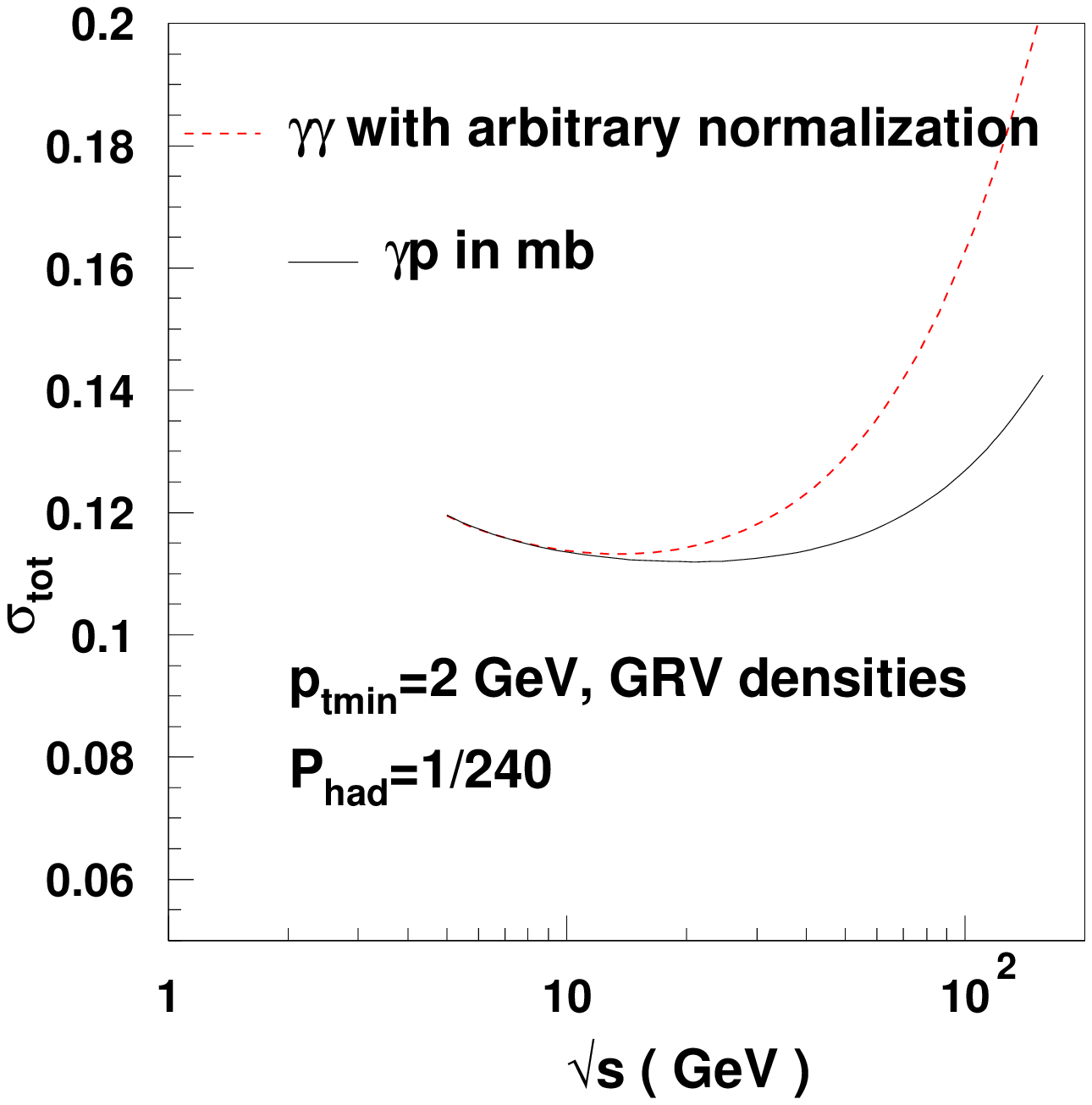,width=10cm}
\caption{Comparison of the energy dependence of the EMM  predictions 
for the total \gamgam\ and \gamp\ cross-sections 
with  $p_{\rm tmin} = 2$ GeV, $P_{\rm had} = 1/240.$}
\label{comp}
\end{center}
\end{figure}
plot in Fig.\ref{comp} the EMM curve for $\gamma p$ and the one 
for $\gamma \gamma$, obtained with the same set of parameters. We have 
normalized the two curves in the low energy region, where they can be 
brought to coincide, thus confirming factorization at low energy. As the 
mini-jet cross-section starts rising, and multiple scattering plays 
a role, then the differences between the impact parameter
distributions, dipole for the proton or monopole for the photon,
become important and the two curves 
do not coincide anymore.

To complete the description of existing models for the $\gamma \gamma$ 
cross-sections, we would like to address the question of whether data 
and models are actually looking at the same quantity; {\it viz.} 
the question of Total versus Inelastic Cross-sections in photon induced 
processes.

For photon induced processes, the 
issue of total cross-section is ill defined both theoretically and 
experimentally. 
In this case,
the $\gamma p$($\gamma \gamma$) cross-sections are extracted from a 
measurement of the $ep(e^+e^-)$ processes. These cross-sections
therefore depend on the acceptance corrections that have to be employed.
These in turn are strongly influenced by the Monte Carlo models to describe
different components of an event. For example, the extraction of 
$\sigma_{\gamma p}^{\rm tot}$ needs understanding of three different kinds 
of events: (i) quasi-elastic process $\gamma p \to V p$, where the proton 
remains intact and the photon gets transformed into a vector meson, (ii)
diffractive where the proton and/or the photon break up but no colour exchange
takes place and (iii) the nondiffractive where both the proton and $\gamma$ 
break up and colour is exchanged between the two. In case of \gamgam\
cross-sections, there are three different kinds of contributions;
(i) the soft interactions modelled by VDM ideas. These have an exponential 
$p_T$ spectrum (ii) the direct interactions of the photon which can be 
estimated using the Quark Parton Model (QPM) and (iii) the resolved 
contributions which rise from the partons in the photon. Let us note that the
first contribution is not to be confused with the nonperturbative part of
the photon structure function. VDM ideas are sometimes used to estimate
this part. These VDM partons also take part in the hard, resolved interactions
which are used in calculating $\sigma_{\gamma p}^{\rm jet}$/
$\sigma_{\gamma \gamma}^{\rm jet}$ in the EMM. So again the extraction
of $\sigma_{\gamgam}$ from the \eplem\ data involves a clear understanding of 
all these types of events. The soft interactions of type (i) are what
can  be losely termed as `elastic' cross-section in this case. Thus we see that
both theoretically and experimentally the ideas about elastic/total 
cross-section are ill defined in the case of photon induced processes.
 
In the earliest applications of the EMM model to the photons~\cite{FLETCHER,
SARC}, 
the inelastic formulation was used. This is correct if the hadrons in the final
state can be defined as all of inelastic origin, i.e. no vector meson decays,
 for
instance. Then, fixing the parameters for the EMM from a fit of the inelastic 
$\gamma p$ cross-section
to the data, one can extrapolate to the $\gamma \gamma$ case, and obtain the
prediction for the inelastic $\gamma \gamma$ cross-section. This
procedure would produce a curve which rises less steeply than the one shown in 
Fig.\ref{figmaria}. On the other hand, as discussed first in
~\cite{aspen} 
and as has been  discussed above, if the data have
correctly included all the diffractive "elastic" processes,  then the
quantity to be compared with the available data should be the `total' 
cross-sections
of eq.~\ref{eik1} and not the inelastic one. If we use  the `total'
cross-section formulation using the eqs.~\ref{eik1}-~\ref{eik3}, use
the known inputs for the parton densities from the photon structure function
measurements and further fix the unknown ad-hoc parameter $p_T^{\rm min}$
of the EMM using the data from $\sigma^{\rm tot}_{\gamma p}$, we then get the
predictions given in Fig.~\ref{figtwo}. If we do the same  using
the inelastic formulation, then we get a curve which at lower energies is
higher, but then rises less fast. 
\begin{figure}[htb]
\begin{center}
\epsfig{file=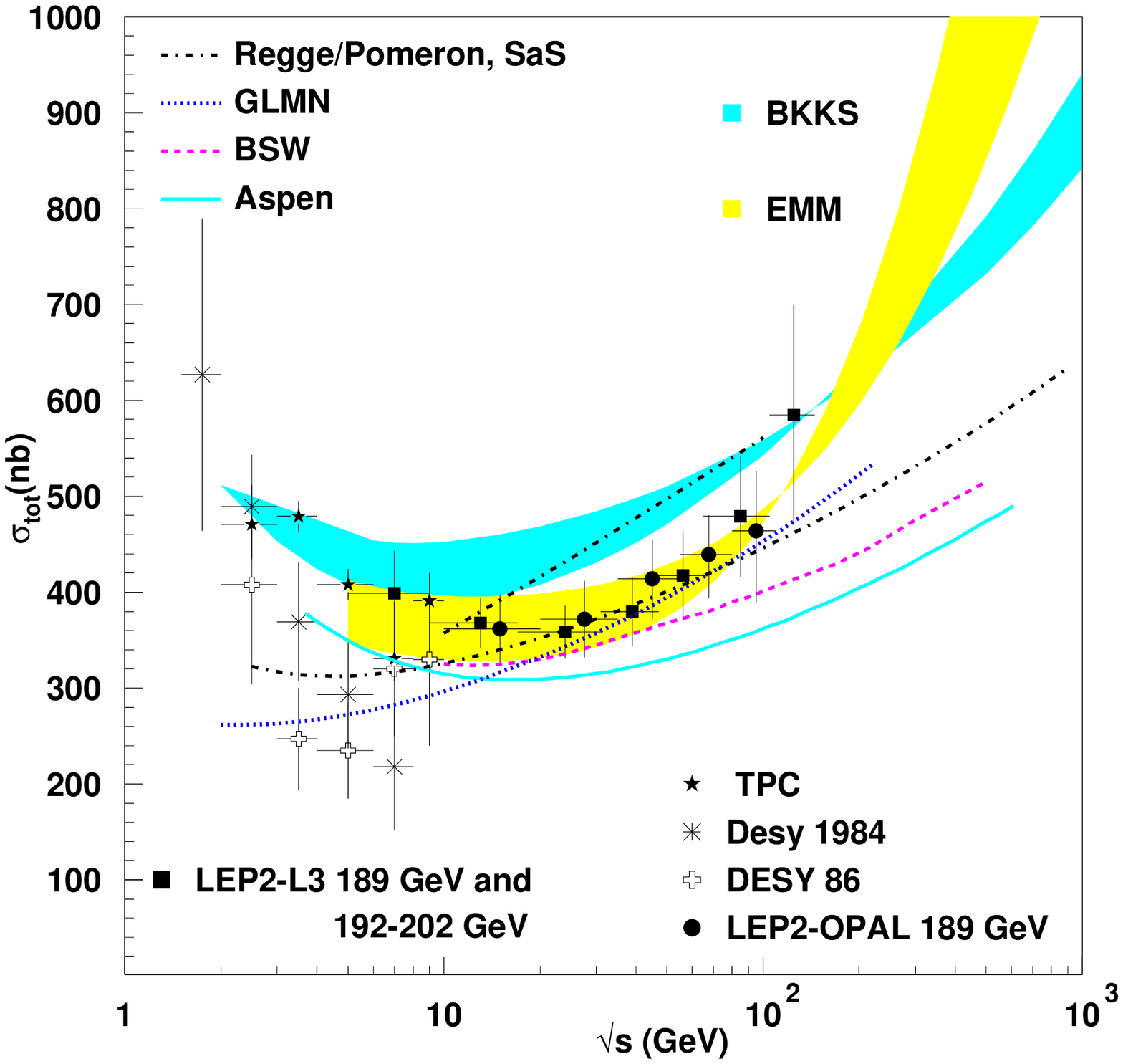,width=10cm}
\caption{The predictions from from factorization models, 
Regge-Pomeron exchange and a QCD structure function models together with 
those
from the EMM and a comparison with present data.}
\label{six}
\end{center}
\end{figure}
In Fig.\ref{six} we have plotted  the
recent data from LEP and from lower energies with the predictions from 
``proton-like'' models, labelled SaS\cite{SAS}, Aspen\cite{aspen},  
BSW\cite{ttwu}, as well as from QCD and Regge inspired
models, like the curve labelled GLMN\cite{glmn} and the band
labelled BKKS\cite{BKS}. 
The band labelled EMM corresponds to the two formulations,
inelastic and total. For the EMM, we have used 
two sets of
representative parameters, both of which are obtained from the $\gamma p$ 
cross-section following the procedure outlined in \cite{emmus}.
\section{Precision necessary}
In this section we show the numerical values corresponding to various
predictions for the total \gamgam\ cross-section and indicate
the precision needed to distinguish among these different
models\cite{ismd99}.
\vspace{-0.7cm}
\begin{table}[htb]
\begin{center}
\caption{Precision required for the measurement of \gamgam\
cross-sections to distinguish between the different `proton' like models}
\vspace{0.3cm}
\begin{tabular}{|c||c|c|c|c||}
\hline 
$\sqrt{s_{\gamma \gamma}} (GeV)$ & Aspen &  BSW & DL & $1 \sigma$ \\ \hline
\hline
 20    & 309 nb & 330 nb & 379 nb &  7\%  \\ \hline
 50    & 330 nb & 368 nb & 430 nb &  11\%   \\ \hline
 100   & 362 nb & 401 nb & 477 nb &  10\%   \\  \hline
 200   & 404 nb & 441 nb & 531 nb &  9\%   \\  \hline
 500   & 474 nb & 515 nb & 612 nb &  8\%   \\  \hline
 700   & 503 nb & 543 nb & 645 nb &  8\%   \\ \hline
\end{tabular}
\label{table1}
\end{center}
\end{table}
In Table ~\ref{table1} we show total $\gamma \gamma$ cross-sections for  
three
models of the `proton-is-like-the-photon' type. The last column shows 
the 1$\sigma$ level precision needed to discriminate between 
Aspen\cite{aspen}
and BSW\cite{ttwu} models. The model labelled DL is obtained from
 Regge/Pomeron exchange with parameters from ref.\cite{DL} and
factorization at the residues. The difference between DL 
and either Aspen or BSW is bigger than between 
Aspen and BSW at each energy value.
A similar table can be drawn for distinguishing between the two minijet 
formulations of Fig.\ref{six} and the BKKS model\cite{BKS}, 
for instance.
\begin{table}[hbt]
\begin{center}
\caption{Precision required for the measurement of \gamgam\ cross-sections
to distinguish between different formulations of the 
EMM and BKKS~\protect\cite{BKS}}
\vspace{0.3cm}
\begin{tabular}{|c||c|c|c|c||}
\hline 
$\sqrt{s_{\gamma \gamma}} (GeV)$ &EMM, Inel,GRS &EMM, Tot,GRV & BKKS& $1 \sigma$ \\ 
& ($p_{\rm tmin}$=1.5 GeV)& ($p_{\rm tmin}$=2 GeV)              & GRV & \\ \hline
\hline
 20    &399  nb & 331 nb      & 408 nb &   2 \%  \\ \hline
 50    &429  nb & 374 nb      & 471 nb &   9\%   \\ \hline
 100   &486  nb & 472 nb      & 543 nb &   11\%   \\  \hline
 200   & 596 nb & 676 nb      & 635 nb&   6\%   \\  \hline
 500   & 850 nb & 1165 nb      & 792 nb &  7  \%   \\  \hline
 700   & 978 nb & 1407 nb     & 860 nb &   13 \%   \\ \hline
\end{tabular}
\label{table2}
\end{center}
\end{table}
The last column in Table~\ref{table2} now gives the percentage 
difference between the two models which bear closest results, 
i.e. EMM with GRS densities and inelastic formulation on the one hand 
and BKKS,  as well as EMM with GRV densities and total formulation on 
the other.

\section{The hadronic backgrounds  at Linear Colliders}
Apart from the above mentioned theoretical interest in studying the 
\gamgam\ cross-sections, a very pragmatic reason is the hadronic backgrounds 
that the beamstrahlung effects might cause at these colliders. One way to
estimate  that is to look at the quantity $\sigma^{\rm jet}_{\gamgam}$ 
defined in eq.~\ref{minjets}. While it is true that only part of the rise
with $\sqrt{s}$ of \siggmjet\ is reflected in  the energy dependence 
of $\sigma^{\rm inel}$, the quantity is still a good measure of the   
messiness caused by the hadronic backgrounds at the NLC due to beamstrahlung.
Here we  give a new parametrisation of the `minijet' 
cross-sections in  \gamgam\ collisions which can be used in estimating the
hadronic backgrounds at the NLC's by folding it with appropriate beamstrahlung
spectra. This supercedes the corresponding parametrisation that was given 
in~\cite{zpcus}.

The 'minijet' cross-section, 
for the two parametrisations GRV~\cite{GRV} and SAS~\cite{SAS} densities,  
is given (in nb) 
\bearr 
\sigma_{minijet}& = &
\left[222 \left( {2 \;{\rm GeV} \over \ptmin}\right)^2 - 161 \left({2\; {\rm GeV} \over \ptmin}\right) 
+ 36.6 \right] \left({\sqrt s\over 50}\right)^{1.23} \label{grv}\\
& =&\left[77.6 \left({2\; {\rm GeV} \over \ptmin}\right)^2 - 45.9 
\left({2\; {\rm GeV} \over \ptmin}\right) + 9.5 \right] \left({\sqrt s \over 50}\right)^{1.17}
\label{eq:sas}
\eearr
by Eqs.~\ref{grv} and \ref{eq:sas} respectively. Here $\sqrt s$ is the 
\gamgam\ c.m. energy in GeV.
Since the dependence on \ptmin\  of \siggmjet\ is extremely strong,
it is essential to fix that.  From our earlier discussions it is clear that
this value will be $\sim 1.5 - 2$ GeV.

Some new theoretical issues that will 
have to be taken in to account in extending these calculations to the higher
energy ($\sqrt s \leq 3-5$ TeV) \eplem\ and \gamgam\ colliders. At these
energies the $x_{\gamma}$ values at which photonic parton densities will 
be sampled will be small ($\gsim 10^{-5}$) and hence saturation effects 
might have to be taken into account. At present, no detailed theoretical 
discussion of the subject is available. These issues might be of relevance 
for the high energy Linear Colliders like CLIC that are beginning to be 
discussed in detail now.

Another aspect of the hadronic backgrounds is also the hadron 
production due to bremsstrahlung photons. This is calculated by convoluting 
the \gamgam\ {\it total } cross-section with the spectrum  of these photons.
This spectrum is given by the  Weizs\"acker Williams(WW) or effective 
photon approximation\cite{WW} which has been very successful in 
translating \gamgam\ cross sections into \eplem\ ones.  There have been many 
disucssions of the improvements on the original WW approximation~\cite{newww}.
The discussion  has also been extended to include the effects due to  
a reduction in the parton content of the photon due to virtuality of 
the photon~\cite{manmev}. The cross-section, including the effects 
due to (anti)tagging of the electron is given by
\begin{equation}
\sigma^{\rm had}_{\eplem} = \int_{zmin}^1 dz_1 
\int_{zmin/z_1}^1 dz_2 \; f_{\gamma/e}(z_1) f_{\gamma/e}(z_2) 
\sigma(\gamgam \to {\rm hadrons}).
\label{ggtoee}
\end{equation}
Here $zmin = s_{min}/s$ where $\sqrt{s}$ is the c.m. energy of the \eplem\  
collider. The WW spectrum used is given by
\be
f_{\gamma/e} (z) = {{\alpha_{\rm em}}\over {2 \pi z}} 
\left[ (1 + (1-z)^2) \ln {{P^2_{max}}\over {P^2_{min}}} -2(1-z) \right],
\label{wwtag}
\ee
where $$
P^2_{max} = s/2*(1-\cos \theta_{tag}) (1-z), P^2_{min}= m_e^2 {z^2 \over (1-z)}.
$$
Here, using $\theta_{tag}$  the maximal scattering angle for the outgoing 
electron, we have taken antitagging into account and have accounted for the 
suppression of the photonic parton densities due to virtuality following
ref.~\cite{zpcus}. 

The antitagging conditions for different beam energies of the  $e^+/e^-$
are modelled using those used at TRISTAN/LEP-1/LEP-II as follows:
\bearr
\theta_{\rm tag} &=&  0.056 , E_{min}^e = 0.25 E_{\rm beam} 
\;\;{\rm for}\;\;
50 < \sqrt{s} < 90 {\rm GeV} \nonumber \\
\theta_{\rm tag} & = & 0.024 ,  E_{min}^e = 0.40 E_{\rm beam}
\;\;{\rm for}\;\; 90 < \sqrt{s} < 200 {\rm GeV} \nonumber\\
\theta_{\rm tag} & = & 0.028 ,  E_{min}^e = 0.25 E_{\rm beam} 
\;\;{\rm for}\;\; 200 < \sqrt{s} < 400 {\rm GeV} 
\nonumber\\
\theta_{\rm tag} & = & 0.025,  E_{min}^e = 0.20 E_{\rm beam}
\;\;{\rm for} \;\; \sqrt{s} > 400 {\rm GeV}. \nonumber\\
\eearr
\begin{figure}
\begin{center}
\epsfig{file=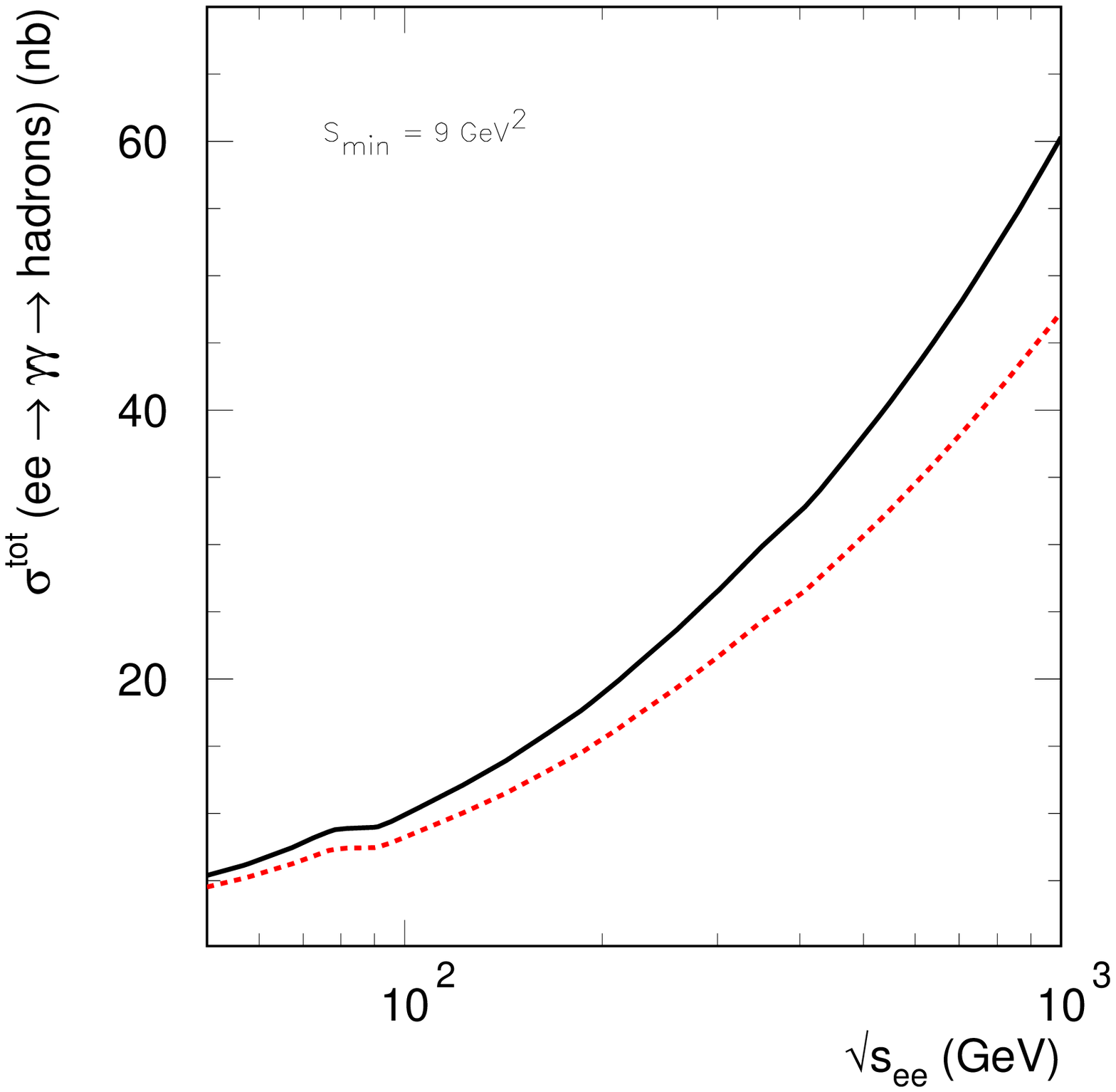,width=8cm}
\caption{Cross-sections for hadron production  due to \gamgam\
interactions in  \eplem\  reactions.}
\label{fig4}
\end{center}
\end{figure}  
In Fig.~\ref{fig4}, we have shown the cross-section 
as a function of $\sqrt{s}$ of the  \eplem\ machine.  The top curve 
corresponds to the prediction  for $\sigma_{\gamgam}$ of the EMM model in the 
inelastic formulation  and the lower curve corresponds to the prediction of
the model~\cite{aspen} for the same. It is to be noted that the difference 
of about factor 2 (say) at $\rts = 700$  GeV, is reduced to about $30\%$
after convolution with the bremsstrahlung spectrum. In this figure we have used 
$s_{min} = 9 \ GeV^2$, to be consistent with the \ptmin = 1.5 GeV used
in the EMM prediction. If we naively extrapolate the predictions
to $s = 1 \ GeV^2$ and thus integrate upto $s_{\rm min} = 1\ GeV^2$ the 
hadron production cross-sections go up by  about a factor 2.  Note also that 
the reduction in the photon spectrum due to the anti-tagging condition 
causes a reduction of about 40\% at the highest end.

\section{Acknowledgments}
We have enjoyed discussions with A. Grau and O. Panella and are grateful
to M. Kienzle for discussions about the recent data and  to A. de Roeck 
for suggestions concerning the precision issue at Linear Colliders. One of us 
is indebted to  M. Block for discussions on the factorization question 
for total cross-sections. This work was supported in part through TMR98-0169.


\begin{thebibliography}{99}
\bibitem{therise} 
D.Cline, F.Halzen and J. Luthe, Phys. Rev. Lett. {\bf 31} (1973) 491.
\bibitem{lcbkgd}
M. Drees and R.M. Godbole, Phys. Rev. Lett. {\bf 67} (1991) 1189;
P. Chen ,T. Barklow and M.E. Peskin, Phys. Rev. {\bf D49} (1994) 3209,
R.M. Godbole, hep-ph/9807379,  Proceedings of the Workshop on{\it
Quantum Aspects of Beam Physics, Jan. 5 1998 - Jan. 9 1998, Monterey, U.S.A.},
404-416, Ed. P. Chen, World  Scientific, 1999. 
\bibitem{review} For a review, see for example, 
M. Drees and R.M. Godbole, Journal of Phys. {\bf G 21} (1995) 1559.
\bibitem{emmus} A. Corsetti, R.M. Godbole and  G. Pancheri, 
Phys.Lett. {\bf B435} (1998) 441.
\bibitem{OPAL}OPAL Collaboration. F. Waeckerle,
{\it Multiparticle Dynamics 1997}, Nucl. Phys. Proc. Suppl. {\bf B71}, 
(1999) 381, Eds. G. Capon, V. Khoze, G. Pancheri and A. Sansoni;
Stefan S\"oldner-Rembold, hep-ex/9810011, To appear in  the proceedings
of the {\it ICHEP'98}, Vancouver, July 1998. 
G. Abbiendi et al.,Eur.Phys.J.C14 (2000) 199. 
\bibitem{L3} L3 Collaboration,
Paper 519 submitted to {\it ICHEP'98}, Vancouver, July 1998.
M. Acciari et al., Phys. Lett. {\bf B 408} (1997) 450; L3 Collaboration,
A. Csilling, Nucl.Phys.Proc.Suppl. {\bf B82} (2000) 239.
\bibitem{L3osaka}L3 Collaboration,
L3 Note 2548, Submitted to the OSAKA Conference.
\bibitem{eikminijets}
L. Durand and H. Pi, Phys. Rev. Lett. {\bf 58} (1987) 58.
A. Capella, J. Kwiecinsky, J. Tran Thanh, Phys.\ Rev.\ Lett.\ 58 (1987) 2015.
M.M. Block, F. Halzen, B. Margolis, Phys. Rev. {\bf  D 45} (1992) 839.
\bibitem{minijet}
A. Capella and J. Tran Thanh Van, Z. Phys. {\bf C 23} (1984)168.
G. Pancheri and C. Rubbia, Nucl. Phys. {\bf A 418} (1984) 117c.
T.Gaisser and F.Halzen, Phys. Rev. Lett. {\bf 54}  (1985) 1754.
P. l`Heureux, B. Margolis and P. Valin, Phys. Rev. {\bf D 32} (1985) 1681.
G.Pancheri and Y.N.Srivastava,  Phys. Lett. {\bf B 158} (1986) 402.
\bibitem{GRV}
M. Gl\"uck, E. Reya and A. Vogt, Zeit. Physik {\bf C 67} (1994) 433 .
M. Gl\"uck, E. Reya and A. Vogt, Phys. Rev. {\bf D 46} (1992) 1973.
\bibitem{treleani}
 D. Treleani and L. Ametller, Int.\ Jou.\ Mod.\ Phys.\ {\bf A 3} (1988) 521 
\bibitem{bn}  A. Grau, G. Pancheri and Y.N. Srivastava,
Phys.Rev. {\bf D60} (1999) 114020.
\bibitem{ladinsky}
J.C. Collins and G.A. Ladinsky, Phys. Rev. {\bf D 43} (1991) 2847.
\bibitem{FLETCHER}
R.S. Fletcher , T.K. Gaisser and F.Halzen, Phys. Rev. {\bf D 45} (1992) 377; 
erratum Phys. Rev. {\bf D 45} (1992) 3279.
\bibitem{SARC}
K. Honjo,  L. Durand, R. Gandhi, H. Pi and I. Sarcevic, Phys. Rev. {\bf D 48} 
(1993) 1048.
\bibitem{lundus} R.M. Godbole and G. Pancheri, Proceedings of the
{\it LUND workshop on photon interactions and photon structure}, 
Aug. 1998, 217-227, Eds. T. Sjostrand and J. Jarsklog, e-print Archive: 
hep-ph/9903331.
\bibitem{ZEUS}
M. Derrick et al., ZEUS coll., Phys. Lett. {\bf B 354} (1995) 163.
\bibitem{photon99} R.M. Godbole, A. Grau and  G. Pancheri.
Nucl.Phys.Proc.Suppl.{\bf B82} (2000), hep-ph/9908220. 
\bibitem{SAS}G. Schuler and T. Sj\"ostrand, Zeit. Physik {\bf C 68} (1995)
607; Phys. Lett. {\bf B 376} (1996) 193.
\bibitem{GRS} M. Gl\"uck, E. Reya and I. Schienbein, Phys.Rev.D60:054019,
1999, Erratum-ibid.D62:019902,2000.
\bibitem{HERAZ} ZEUS Collaboration, Phys. Lett. {\bf B 293} (1992), 465; 
Zeit. Phys. {\bf C 63} (1994) 391.
\bibitem{HERAH1} H1 Collaboration, 
Zeit. Phys. {\bf C69} (1995) 27.
\bibitem{DIS}  J. Breitweg et al., ZEUS coll., {\bf DESY-00-071}, e-print 
Archive: hep-ex/0005018.
\bibitem{aspen} M.M. Block, E.M. Gregores, F. Halzen and  G. Pancheri,
 Phys.Rev.{\bf D58} (1998) 17503; 
M.M. Block, E.M. Gregores, F. Halzen and G. Pancheri,
Phys.Rev. {\bf D60} (1999) 54024.
\bibitem{ttwu} C. Bourelly, J. Soffer and T.T. Wu, Mod.Phys.Lett. {\bf A15}
(2000) 9. 
\bibitem{DL} A. Donnachie  and P.V. Landshoff, 
Phys. Lett.{\bf B 296} (1992)  227.
\bibitem{BKS} B. Badelek, M. Krawczyk, J. Kwiecinski and  A.M. Stasto. 
e-Print Archive: hep-ph/0001161.
\bibitem{glmn} E. Gotsman, E. Levin, U. Maor, E. Naftali, 
Eur.Phys.J. C14 (2000) 511,
   hep-ph/0001080. 
\bibitem{ismd99}R. Godbole, A. Grau and G. Pancheri, 
{\it QCD and Multiparticle Production}, page 424. Proceedings of the XXIX 
International Symposium on Multiparticle
Dynamics, Brown U., Aug. 2000.
 Eds. I. Sarcevic and C-I Tang. World Scientific 2000. e-print Archive:
hep-ph/9912395.
\bibitem{zpcus} M. Drees and R.M. Godbole, Zeit. Phys. C {\bf 59} (1993) 591.
\bibitem{WW} C.F. v. Weizs\"acker, Z. Phys. {\bf 88}, 612 (1934); 
E.J. Williams, Phys. Rev.  {\bf 45} 729.
\bibitem{newww} For a recent discussion, see for example,
S. Frixione, M.L. Mangano, P. Nason and G. Ridolfi, Phys. Lett. {\bf B319},
339 (1993).
\bibitem{manmev} M. Drees and R.M. Godbole, Phys. Rev. D {\bf 50} (1994) 3124.
\end{thebibliography}
\end{document}